\documentclass[a4paper]{jpconf}
\usepackage{graphicx}
\usepackage{slashed}
\usepackage{subfig}
\usepackage{float}

\newcommand{\neut}[1]{\widetilde{\chi}^0_{#1}}
\newcommand{\cha}[1]{\widetilde{\chi}^\pm_{#1}}
\newcommand{\met}{\ensuremath{\slashed E_T}}
\newcommand{\muco}{\ensuremath{\mu_{\rm col}}}

\begin{document}

\pagestyle{plain}
\setlength{\footskip}{30pt}

\begin{flushright}
KIAS-P17026\\
\end{flushright}

\title{Light exotic Higgs bosons at the LHC}
\thispagestyle{plain}
\setlength{\footskip}{30pt}

\author{Shoaib Munir}

\address{School of Physics, Korea Institute for Advanced Study,
Seoul 130-722, Republic of Korea}

\ead{smunir@kias.re.kr}

\begin{abstract}
Most models of new physics contain extended Higgs sectors with multiple Higgs bosons. The observation of an additional Higgs boson, besides the $\sim 125$\ GeV `$h_{\rm obs}$', will thus serve as an irrefutable evidence of physics beyond the Standard Model (SM). However, even when fairly light, these additional Higgs bosons may have escaped detection at the Large Electron-Positron (LEP) collider, the Tevatron and the Large Hadron Collider (LHC) hitherto, owing to their highly reduced couplings to the SM particles. Therefore, in addition to the searches based on the conventional production processes of these Higgs bosons, such as gluon or vector boson fusion, possible new search modes need to be exploited at collider experiments in order to establish their signatures. We investigate here the phenomenology of pseudoscalars, with masses ranging from $\mathcal{O}$(1)\ GeV to about 150\ GeV, in the Next-to-Minimal Supersymmetric SM (NMSSM) and the Type-I 2-Higgs Doublet Model (2HDM) in some such atypical search channels at the LHC Run-II.
\end{abstract}

\section{Introduction}

In any new physics model with an extended Higgs sector, (at least) one of the neutral Higgs bosons should have a mass and signal rates in its various decay channels consistent with those of the $h_{\rm obs}$ discovered at the LHC [1,\,2]. This requirement, coupled with that of  the satisfaction of constraints coming from precision electroweak (EW) and $b$-physics experiments, almost always leads to the other two of the three neutral Higgs bosons being rather heavy in the minimal realisation of Supersymmetry (SUSY). However, in the NMSSM, where the augmentation by a singlet scalar superfield results in a total of five neutral Higgs states, scalars $H_{1-3}$  (ordered in terms of increasing mass) and pseudoscalars $A_{1,2}$, the singlet-like states can be fairly light (for a review, see [3]). A light $A_1$ can prevent the over-closure of the universe when the lightest neutralino, $\chi_1^0$, which is a crucial dark matter (DM) candidate, has a mass so low that neither the $Z$ boson nor the $h_{\rm obs}$ can contribute sufficienty to its annihilation.

Without invoking SUSY, one can simply introduce a second Higgs doublet in the SM, with a $Z_2$-symmetry preventing the dangerous flavor changing neutral currents (for a review, see [4]). In a Type-I 2HDM (2HDM-I), this $Z_2$ symmetry is imposed in such a way that all the fermions couple only to one of the two Higgs doublets. The physical masses of the three neutral Higgs bosons can be taken to be the input parameters in this model. One can therefore assign any masses to the scalars $h$ and $H$ (with $m_h < m_H$) and the pseudoscalar $A$ in order to study the phenomenological implications for different values of the other free parameters.

We analyse some of the potential discovery channels of a light pseudoscalar at the LHC Run-II in these models. In the NMSSM we consider the production of $A_1A_1$ and $A_1Z$ pairs in the decays of the heavier CP-even scalars [5] and of very light $A_1$ and DM in the decays of the heavier neutralinos [6]. In the 2HDM-I we discuss the electroweak (EW) production of an $hA$ pair with a combined mass less than that of the $Z$ boson [7].

\section{Analysis methodology}
For the first of our two NMSSM analyses, in order to perform a fast and efficient numerical scanning of the parameter space for obtaining $A_1$ solutions spanning a wide range of masses, we adopted a model version with partial universality. In this version, unified parameters $m_0$, $m_{1/2}$ and $A_0$, corresponding to the scalar masses, gaugino masses and trilinear couplings, respectively, are input at the grand unification scale. The Higgs sector soft trilinear parameters, $A_\lambda$ and $A_\kappa$, though not unified with $A_0$, are also input at the same high scale. In contrast, the dimensionless Yukawa couplings $\lambda$ and $\kappa$, and the parameters $\mu_{\rm eff}\equiv \lambda s$ and $\tan\beta \equiv v_u/v_d$, with $v_u$ and $v_d$ being the vacuum expectation values of the two Higgs doublets and $s$ of the singlet, are defined at the EW scale. These input parameters were fed into the NMSSMTools [8] program to calculate the physical masses, couplings and branching ratios (BRs) of all the Higgs bosons. For the 2HDM-I, the Higgs boson couplings and BRs were obtained for the scanned ranges of the input parameters using the 2HDMC public code [9]. 

During the scanning process, each point was first subjected to the basic theoretical conditions like unitarity, perturbativity and vacuum stability. For every point fulfilling these conditions, we further required the $H_2$ ($H$) in the NMSSM (2HDM-I) to have its mass and signal strengths, calculated with HiggsSignals [10], consistent with the latest available measurements for the $h_{\rm obs}$ from the LHC (see, e.g., [11]). A successful model point was then tested for consistency of the remaining scalar, pseudoscalar and charged Higgs bosons with the exclusion limits from collider searches, using the HiggsBounds program [12]. It was also required to satisfy the constraints on the most important $b$-physics observables, the predicted values for which were calculated for a point in each of the models using the SuperIso [13] program. In addition, for the 2HDM-I points, the values of the oblique parameters $S$, $T$ and $U$, calculated by 2HDMC, were checked against exclusion limits from the experimental measurements [14]. In the case of the NMSSM, on the other hand, a point was rejected if the relic abundance due to the $\chi_1^0$ was computed by MicrOmegas [15] to be larger than the measurement by the PLANCK telescope [16]. The scanned ranges of the free parameters in the two models are given in table \ref{tab:params}. 

 \begin{table}[h]
\begin{center}
\caption{Scanned ranges of the free parameters in the NMSSM (a) and the 2HDM-I (b).}
\label{tab:params}
\begin{tabular}{cc}
\lineup
\subfloat[]{%
\begin{tabular}{*{2}{l}}
\hline
NMSSM parameter & Scanned range  \\
\hline
$m_0$\,(GeV) 	& \m\0200 -- 4000 \\
$m_{1/2}$\,(GeV)  & \m\0100 -- 2000	\\
$A_0$\, (GeV)  & $-5000$ -- 0\\
$\tan\beta$ 		& \m\0\0\01 -- 40 \\
$\lambda$ 		& \0\00.01 -- 0.7 \\
$\kappa$ 		& \0\00.01 -- 0.7 \\
$\mu_{\rm eff}$\,(GeV) 	& \m\0100 -- 2000 \\
$A_\lambda$\,(GeV)  	& $-2000$ -- 2000 \\
$A_\kappa$\,(GeV)  	& $-2000$ -- 2000\\
\hline
\end{tabular}
}
&
\lineup
\subfloat[]{%
\begin{tabular}{ll}
\hline
2HDM-I parameter & \0Scanned range \\
\hline
\hline
		$m_h$\ (GeV) & \0\0\0~10 -- $2M_Z/3$ \\
		$m_A$\ (GeV) & \0$m_h/2$ -- ($M_Z-m_h$) \\
		$m_{H^\pm}$ (GeV) & \0\0\0~90 -- 150 \\
		$\sin(\beta-\alpha)$ & ~$-0.25$ -- 0 \\
		$m_{12}^2$ GeV$^2$ & \0\0\0\0~0 -- $m_A^2\sin\beta\cos\beta$ \\
		$\tan\beta$ & \0{\normalsize $\frac{(-0.95~\textrm{--}~-1.1)}{\sin(\beta -\alpha)}$} \\
\hline
	\end{tabular}
}
\end{tabular}
\end{center}
\end{table}

The second NMSSM analysis included here is dedicated to a very specific scenario in which the $A_1$ and the $\chi_1^0$ both have masses $\mathcal{O}$(1)\ GeV. Thus only a couple of benchmark points (BPs) corresponding to the general NMSSM, with all the input parameters lying at the EW scale, will be discussed. For both these BPs,  the role of the $h_{\rm obs}$ is once again played by the $H_2$, and all the relevant constraints noted above are satisfied.

\section{Light pseudoscalars in the NMSSM}
In the NMSSM, the tree-level mass-squared of the $A_1$ is written (assuming negligible singlet-doublet mixing) as
\begin{equation}
\label{eq:ma2}
m_{A_1}^2 \simeq \frac{A_\lambda}{\sqrt{2}s}  v^2 \lambda\sin 2\beta + \kappa (2  v^2 \lambda \sin 2\beta - 3sA_\kappa)\,,
\end{equation}
 where $v \equiv \sqrt{v_u^2 + v_d^2} \simeq 246$\ GeV. Thus, $m_{A_1}$ can be varied with more freedom compared to the mass of the singlet-like scalar, which is also indirectly constrained by the $h_{\rm obs}$ measurements due to the relatively stronger singlet-doublet mixing. By adjusting the trilinear couplings $A_\lambda$ and $A_\kappa$, $A_1$ can take a broad range of values without being in conflict with the experimental data. 
 
\subsection{Heavy Higgs boson decays}
For this analysis, we restricted ourselves to $m_{A_1} < 150$\ GeV, so that the production cross section for the heavier Higgs bosons that decay into it did not get too suppressed kinematically. After performing the parameter space scan to find model points satisfying all the imposed conditions, we carried out a dedicated
signal (S)-to-background (B) analysis for the LHC with $\sqrt{s}=14$\ TeV. We first calculated the gluon-fusion production cross section for each of the $H_{1-3}$ using the public program SusHi [17]. The backgrounds coming from the $pp\to 4b$,
$pp\to 2b2\tau$, $pp\to 4\tau$, $pp\to Z 2b$ and $pp\to Z2\tau$ processes were computed using MadGraph5\_aMC@NLO [18]. The hadronization and fragmentation of the signals and backgrounds was then done using Pythia 8.180 [19] interfaced with FastJet [20]. 

We used the two most dominant decay channels of the $A_1$, namely $b\bar{b}$ and $\tau^+\tau^-$. In the case of the $b\bar{b}$ decay, we employed also the  jet substructure method [21], which gives enhanced sensitivity for larger masses of the decaying Higgs bosons, by assuming one fat jet from boosted $b$-quarks instead of two single $b$-jets. For three representative values of the accumulated luminosity at the LHC, $\mathcal{L}=30$/fb, 300/fb and 3000/fb, we then estimated the signal cross sections which would give the statistical significance, $S/\sqrt{B}$, greater than 5 for a given mass of $A_1$ in each of the various final state combinations. 

\begin{figure}[tbp]
\subfloat[]{%
\includegraphics*[width=8cm]{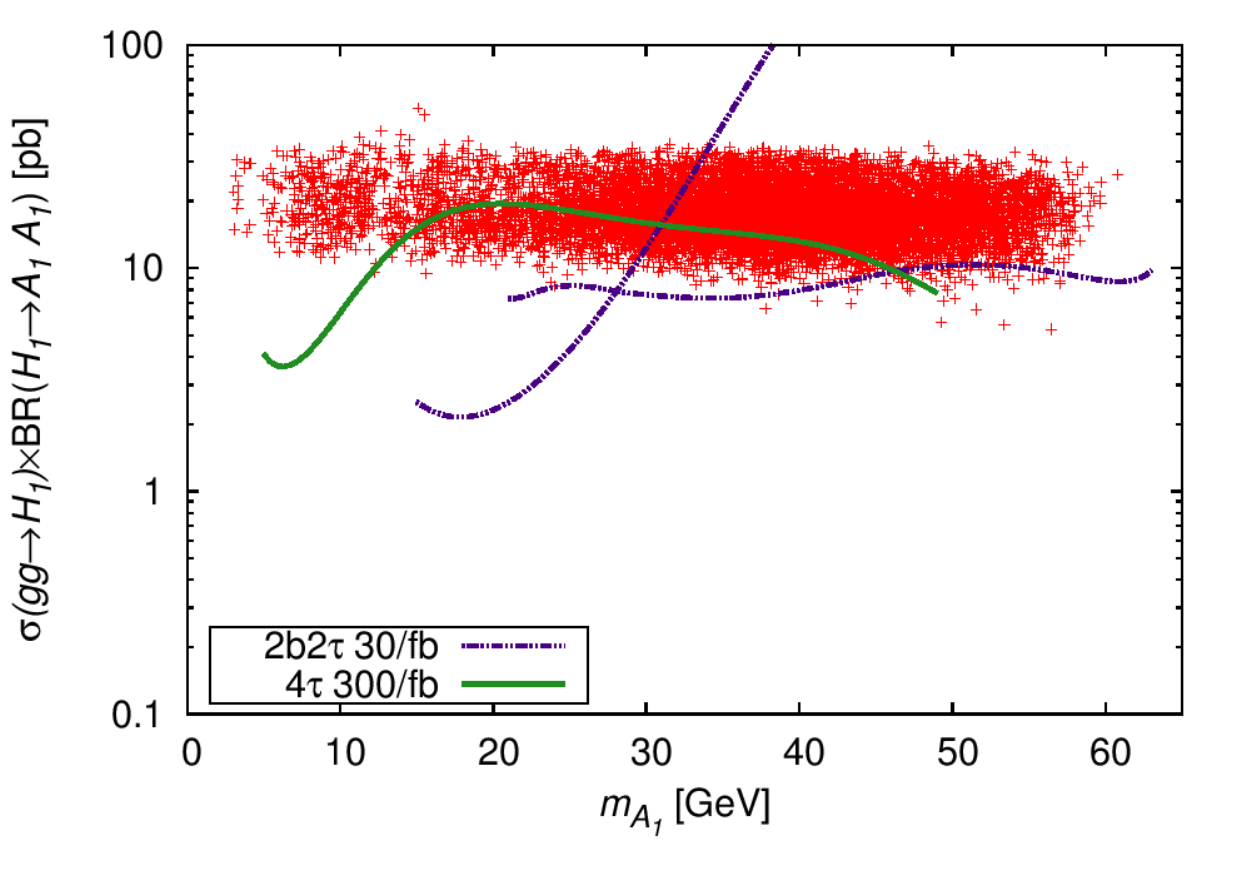}
}
\hspace{0.2cm}%
\subfloat[]{%
\includegraphics*[width=8cm]{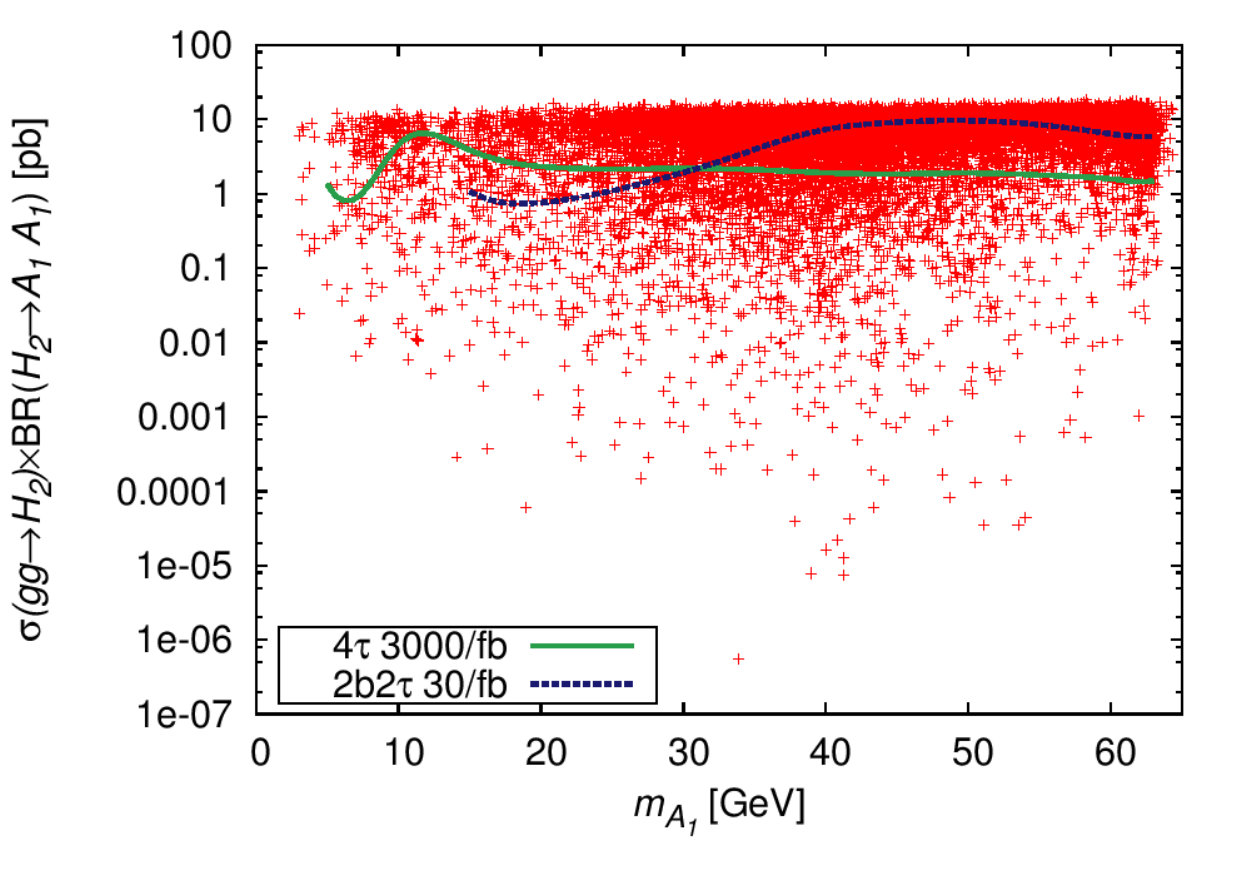}
}
\caption{Cross sections for (a) the $gg \to H_1 \to  A_1A_1$ process, and (b) the $gg \to H_2 \to  A_1 A_1$ process, as functions of $m_{A_1}$, for points obtained from the NMSSM scan. Taken from [5].}
\label{fig:H2toA1}
\end{figure}

In figure \ref{fig:H2toA1}(a) we show the cross section for the $H_1\rightarrow A_1A_1$ process. Also shown, in this figure and the subsequent ones, are the sensitivity curve(s), which assume BR$(A1 \to b\bar{b})=0.9$ for each $b\bar{b}$ pair and BR$(A1\to \tau^+\tau^-)=0.1$ for each $\tau^+\tau^-$ pair, corresponding to the best final state combination for probing the given process. Here these curves are for the $2b2\tau$ final state (one corresponding to two single $b$-jets and the other, showing an enhanced sensitivity for smaller $m_{A_1}$, to one fat jet) at $\mathcal{L}=30$/fb, and for the $4\tau$ final state at $\mathcal{L}=3000$/fb. In the frame (b) we show the $H_2 \rightarrow A_1A_1$ cross section. Note that in the case of the $H_2$, the possibility to reconstruct its mass (125\ GeV) provides an important kinematical handle. We see that the $2b2\tau$ final state  can be probed at the LHC with $\mathcal{L}$ as low as 30/fb, owing to the use of the jet substructure method, despite the fact that the $H_2\to A_1 A_1$ decay is tightly constrained by the $h_{\rm obs}$ signal rate measurements at the LHC. However, the maximum $A_1$ mass that can be accessible in this channel is $m_{H_2}/2 \sim 62.5$\ GeV. Evidently, only smaller values of $m_{A_1}$ might be accessible in the $H_2/H_1\to ZA$ decay channels. We therefore ignore these channels and turn to the $H_3$ for the production of heavier $A_1$.    

\begin{figure}[tbp]
\subfloat[]{%
\includegraphics*[width=8cm]{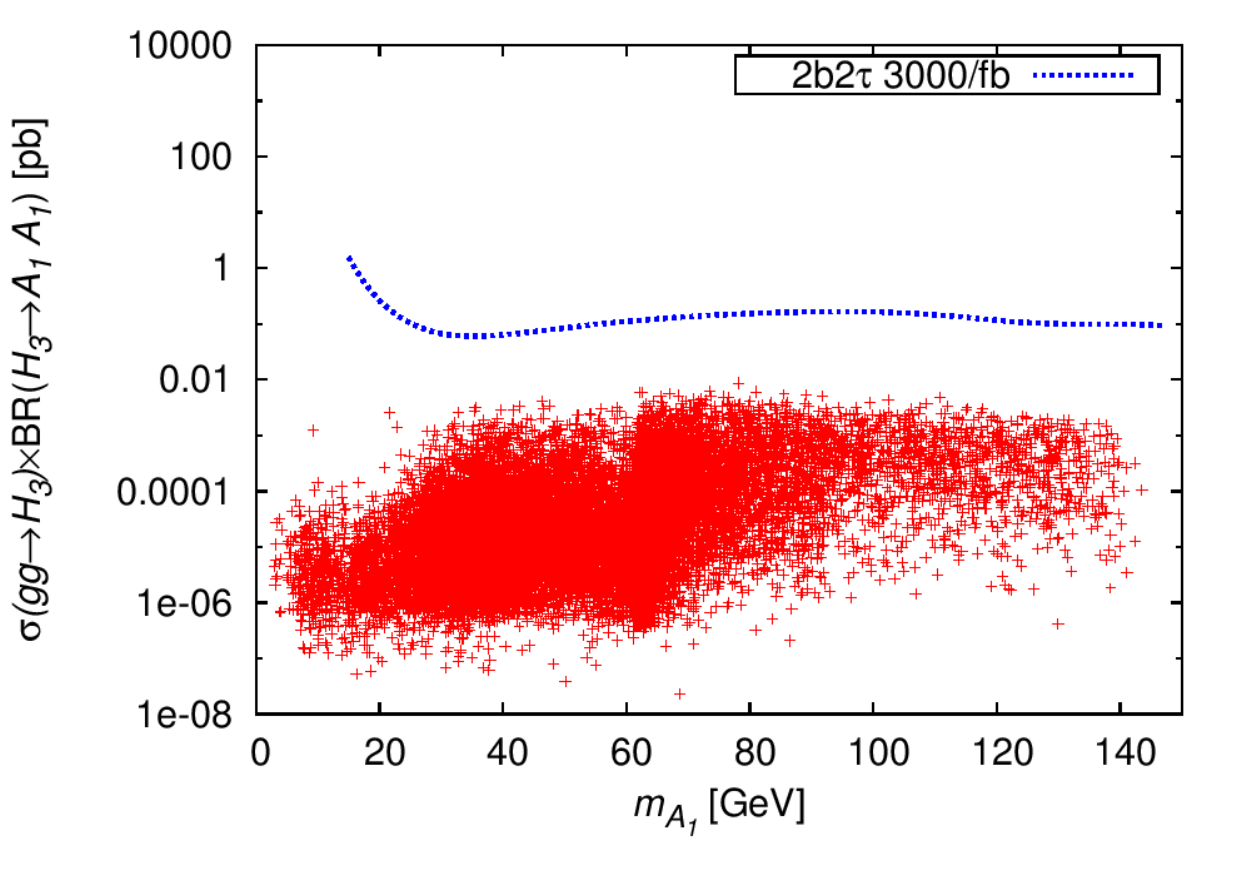}
}
\hspace{0.2cm}%
\subfloat[]{%
\includegraphics*[width=8cm]{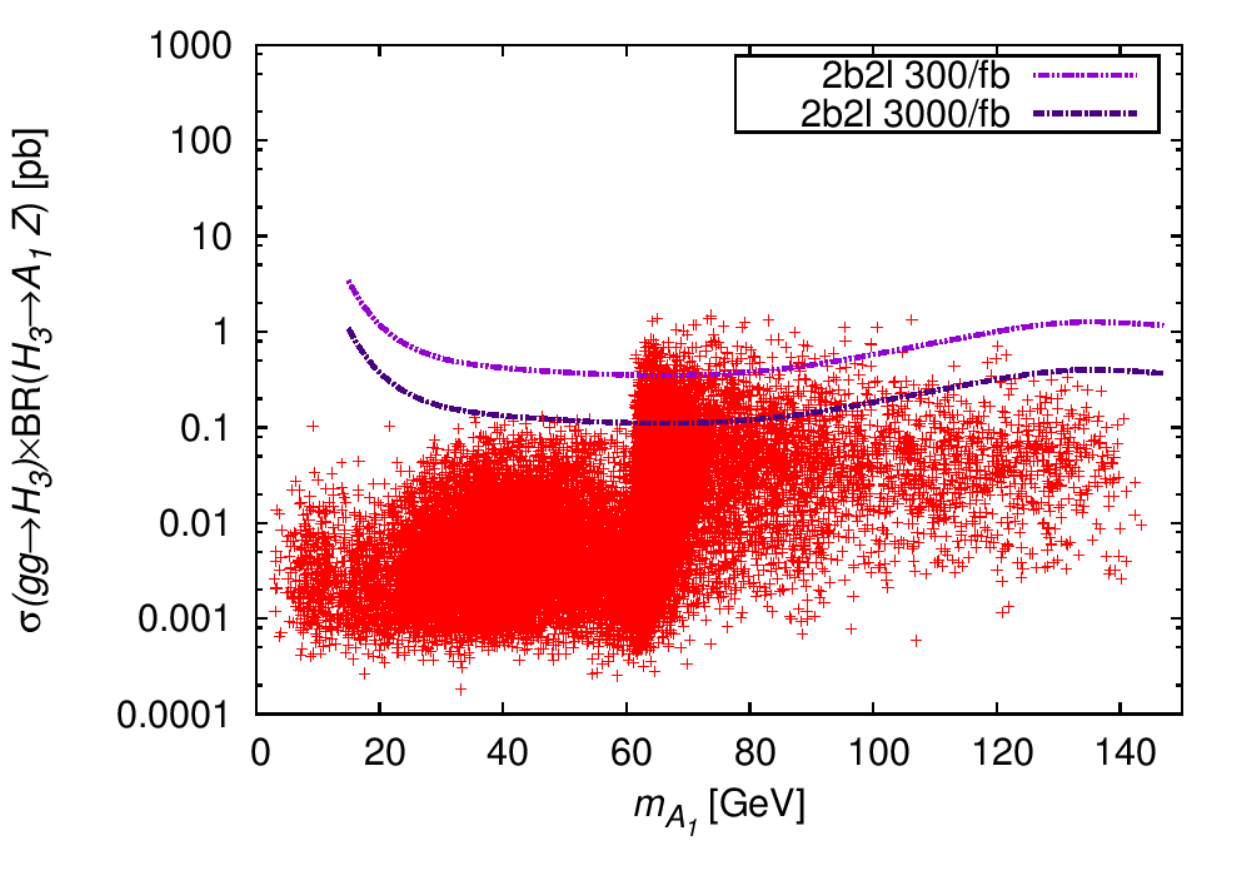}
}
\caption{Cross sections for (a) the $gg \to H_3 \to  A_1A_1$ process, and (b) the $gg \to H_3 \to  A_1 Z$ process, as functions of $m_{A_1}$, for points obtained from the NMSSM scan. Taken from [5].}
\label{fig:H3toA1}
\end{figure}

Figure \ref{fig:H3toA1}(a) shows that the $H_{3} \rightarrow A_1A_1$ channel does not carry any promise. This is due to the fact that for such high masses of $H_3$ ($\geq 400$\ GeV) the production cross section gets diminished and, at the same time, other decays of $H_3$ dominate over this channel. The sensitivity curve in the figure corresponds to the $2b2\tau$ final state for $\mathcal{L}=3000$/fb. Conversely, as seen in the frame (b), for the $H_3\to A_1Z$ channel, with the $Z$ decaying into $e^+e-$ or $\mu^+\mu^-$ states, a number of points lie above the $2b2\ell$ sensitivity curve for $\mathcal{L}=300$/fb. Again, the use of the fat jet analysis, along with a sizable $H_3A_1Z$ coupling resulting from a significant doublet component in $A_1$, make an $A_1$ lying in the $\sim 60 - 100$\ GeV mass range discoverable in this channel.

\subsection{DM associated production}
Next we focus on the $A_1$ with mass $\mathcal{O}$(1)\ GeV in the NMSSM, which also contains five neutralinos, $\neut{1-5}$, in total. The lightest of these states, given by the linear combination
 \begin{eqnarray}
 \neut{1} = N_{11} \widetilde B^0 + N_{12} \widetilde W_3^0 + N_{13} \widetilde
 H_d^0 + N_{14} \widetilde H_u^0 + N_{15} \widetilde S^0,
 \end{eqnarray}
of the gaugino ($\widetilde B^0,\widetilde W_3^0$), higgsino ($\widetilde H_u^0,\widetilde H_d^0$) and singlino ($\widetilde S^0$) interaction eigenstates, is the DM candidate when $R$-parity is conserved. The presence of the singlino fraction, 
$Z_s=|N_{15}|^2$ in the $\neut{1}$, which is non-existent in the MSSM, leads to some interesting new possibilities in the context of DM phenomenology. A look at the NMSSM neutralino mass matrix reveals that the $[\mathcal{M}_{\widetilde{\chi}^0}]_{55}$ term, corresponding to the singlino eigenstate, is equal to 
$2\kappa s = 2\frac{\kappa \mu_{\rm eff}}{\lambda}$. This implies that the singlino fraction in $\neut{1}$ can be increased by reducing $\kappa$ and/or $\mu_{\rm eff}$ and increasing $\lambda$. Since the mass of $A_1$ also scales with $\kappa s$, as noted above, a light $A_1$ can naturally accompany a light singlino-like $\neut{1}$. This DM can thus undergo sufficient annihilation, via $A_1$ in the $s$-channel, to generate the correct relic abundance of the universe. 

At the LHC, one of the main ways to probe the DM is in the decays of the heavier neutralinos and charginos. In particular, dedicated searches have been performed by both the CMS and ATLAS collaborations [22,\,23] for the $pp \to \neut{2,3} + \cha{1}$ process which is followed by the decays $\neut{2,3}\to Z + \neut{1} \to \ell^+\ell^- + \met$ and $\cha{1}\to W^\pm + \neut{1} \to \ell^\pm + \met$, where \met\ implies missing transverse energy. These searches have already put strong constraints on significant regions of the NMSSM parameter space, since the $Z + \neut{1}$ decay channel is by far the dominant one of $\neut{2,3}$. However, in the scenario with a very light singlino-like DM, the $\neut{2,3} \to A_1 + \neut{1}$ decay channel, although still subdominant, can become sizable for sufficiently
large values of $\lambda$. The reason is that the $\neut{2}$, $\neut{3}$ and $\cha{1}$ are predominantly higgsinos, since $\mu_{\rm eff}$ is much smaller than the gaugino soft masses $M_1$ and $M_2$, in order to maximize the singlino fraction in $\neut{1}$ while minimizing its mass. The issue with the $A_1 + \neut{1}$ decay mode though, is that the main leptonic decay channel, $A_1 \to \mu^+ \mu^-$, is highly suppressed, with its BR never exceeding 9\%. Furthermore, the muons thus produced are highly collinear and hence the isolation of this signal from the background is extremely challenging. 

\begin{figure}[h]
\subfloat[]{%
\includegraphics*[width=7.5cm]{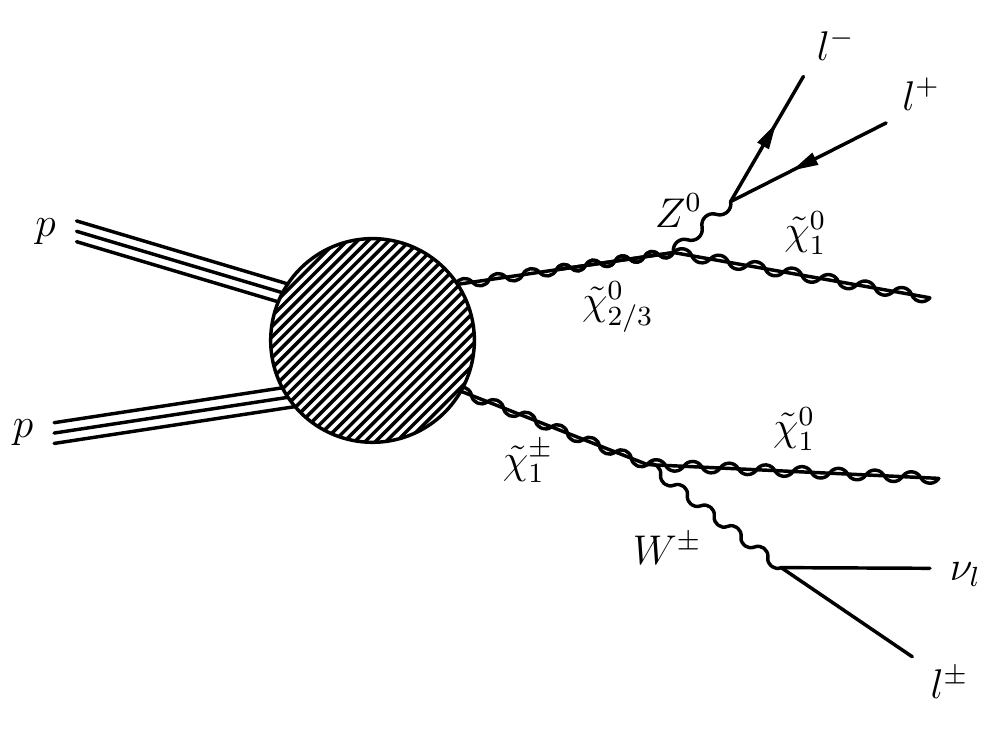}
}
\hspace{0.5cm}%
\subfloat[]{%
\includegraphics*[width=7.5cm]{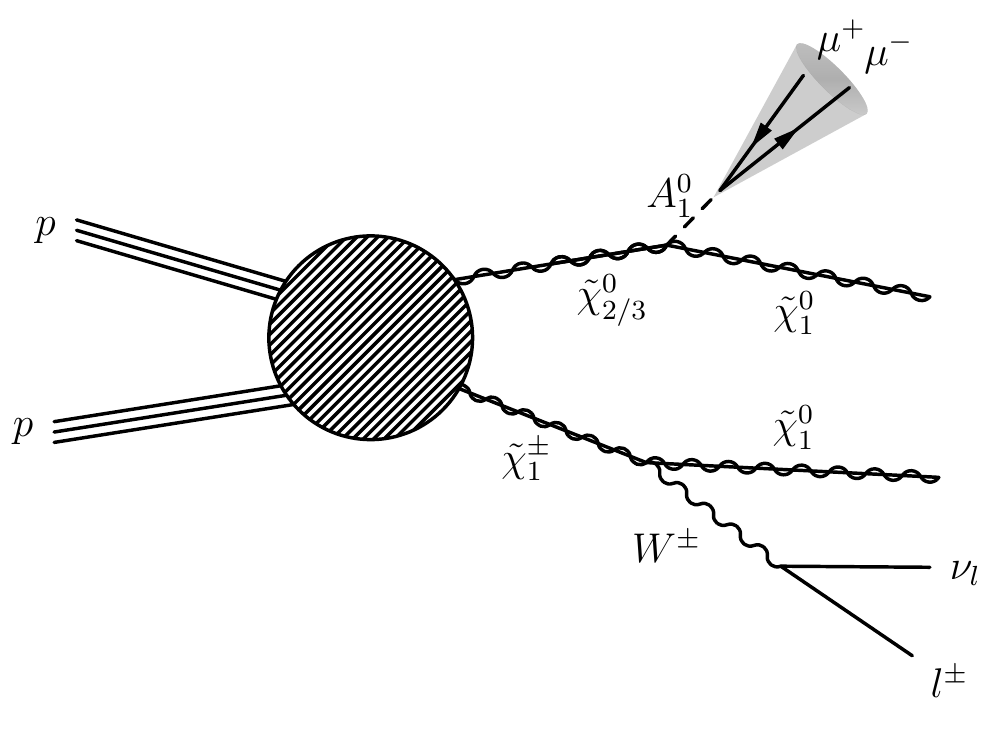}
}
\caption{Diagramatic representation of the processes containing, in the final state, (a) two leptons coming form a $Z$ boson, and (b) two collinear muons coming from an $A_1$.}
\label{fig:processes}
\end{figure}

We show here that, for the NMSSM parameter space points yielding $\mathcal{O}$(1)\ GeV $\neut{1}$ and $A_1$, once the above complication can be overcome, the $A_1 + \neut{1}$ search channel can be more promising than the $Z + \neut{1}$ one at the LHC.  We refer to the former as the $\mu_{\rm col}$ channel and to the latter as the trilepton ($3\ell$) channel. For this purpose, from the NMSSM parameter space, we chose BP1 such that the $\neut{2,3} \to A_1 \neut{1}$ decays are typically suppressed (both having BRs of 0.004), while BP2 has relatively enhanced respective BRs of 0.089 and 0.081. The BRs corresponding to  the $\neut{2,3} \to Z \neut{1}$ decays for both the points are in excess of 60\%, while the $BR(A_1\rightarrow \mu^+\mu^-)$ is 0.039 for BP1 and 0.087 (i.e., near its maximum possible value) for BP2. For these BPs we then performed detector-level analyses of the two processes shown in figure \ref{fig:processes}. 

For the $3\ell$ channel, we first generated parton-level signal and background events at the 14\ TeV LHC for each of the BPs using MadGraph5\_aMC@NLO and passed these to Pythia 6.4 [24] for hadronization. The most dominant irreducible backgrounds for this channel come from the di-boson, tri-boson and $t\bar{t}W/Z$ productions, all of which can have three or more leptons and $\met$ in the final states. 
To obtain the signal and background efficiencies, the ATLAS detector simulation was then performed with DELPHES 3 [25] via the CheckMATE program [26], wherein the six distinct signal regions defined in the ATLAS search [23] have already been implemented. By multiplying the next-to-leading order cross sections for the signal process, calculated using Prospino [27], and the backgrounds with an assumed $\mathcal{L} = 300\,{\rm fb}^{-1}$, we also obtained the number of events for both in each of the signal regions.

As for the \muco\ channel, in order to isolate the highly collimated muons, we employed the technique of clustering them together into one object, similar in concept to the construction of a ‘lepton-jet’ [28]. For applying this method, the signal events generated for BP1 and BP2 were passed to Pythia 6.4 for hadronization and subsequently to DELPHES 3 for jet-clustering using Fastjet. Then, the \muco\ object was defined by requiring the transverse momentum, $p_T$, for each muon in the signal to be larger than 10\,GeV, and the cut $m_{\mu\bar \mu}<5$\,GeV was imposed on the invariant mass of the muon pair. It additionally satisfied the condition $I_{\rm sum} < 3 $\ GeV, with $I_{\rm sum}$ being the scalar sum of the transverse momenta of all additional charged tracks, each with $p_T > 0.5$\,GeV, within a cone centered along the momentum vector of \muco\ and satisfying $\Delta R = 0.4$.  The main backgrounds, containing two collinear muons along with a third lepton and \met, include $W(\to \ell^{\pm}v)\gamma^*$, $Z(\to
\ell^+\ell^-)\gamma^*$ and $Wb\bar{b}$ and $Z\gamma^*$. These backgrounds were also generated with Pythia and Fastjet and subjected to certain cuts that maximize the isolation of the signal process from them, as explained in [6]. 

The yield of each of the above analysis methods applied to the two search channels is quantified in terms of $S/B$, which is given in table~\ref{tab:bench} for the two BPs. For the $3\ell$ channel, this $S/B$ corresponds only to the signal region that gives the highest sensitivity, and we note that it is slightly higher than the $S/B$ in the \muco\ channel for the BP1. For the BP2, however, the $\muco$ analysis gives a considerably larger $S/B$ than the $3\ell$ one, which is evidently a consequence of the sizable BR($\neut{2,3} \to A_1 \neut{1}$) and BR($A_1\to \mu^+\mu^-$). Thus dedicated searches in the \muco\ channel may prove very crucial for the discovery of a very light DM in non-minimal supersymmetry at the LHC. Note that while an estimation of the statistical significance would be a more realistic indicator of the strengths of the two signal processes compared to the $S/B$, it is not included here since there is no consistent way of treating the systematic uncertainties. 

\begin{table}[th!]
\caption{Measures of the strengths of the two analyses considered here, along with the masses, in GeV, relevant to them, corresponding to the two selected BPs.}
\label{tab:bench}
\centering
\begin{tabular}{l l l l l l l l l}
\br
BP & $m_{\neut{1}}$  & $m_{\neut{2}}$  & $m_{\neut{3}}$ & $m_{\cha{1}}$ & $m_{A_1}$ & $m_{H_2}$  & $S/B$ ($3\ell$)  & $S/B$ ($\muco$) \\
\mr
1 & 1.00 & 189.1  & $-201.7$ & 195.0 & 2.18 & 124.1 & 0.591 & 0.42 \\
2 & 1.41 & 170.1 & $-182.3$  & 167.7 & 2.99 & 125.8 & 0.436 & 15\\
\br
\end{tabular}
\end{table}

\section{Scalar-pseudoscalar pair-production in the Type-I 2HDM}

The Landau-Yang theorem [29,\,30] prevents the contribution of an on-shell $Z$ boson to the gluon-initiated production of a $hA$ pair when the sum of their masses is smaller than $m_Z$. The $q\bar{q}$-initiated process, however, does not suffer from this limitation, and the cross section for $hA$ pair-production can therefore get considerably enhanced due to a resonant $Z$ boson in the $s$-channel. Our analysis of the Type-I 2HDM aimed at exploring this possibility, and hence the parameter space scan for this model also observed the condition $m_h + m_A < m_Z$. 

\begin{figure}[h]
\subfloat[]{%
\includegraphics*[width=7.8cm]{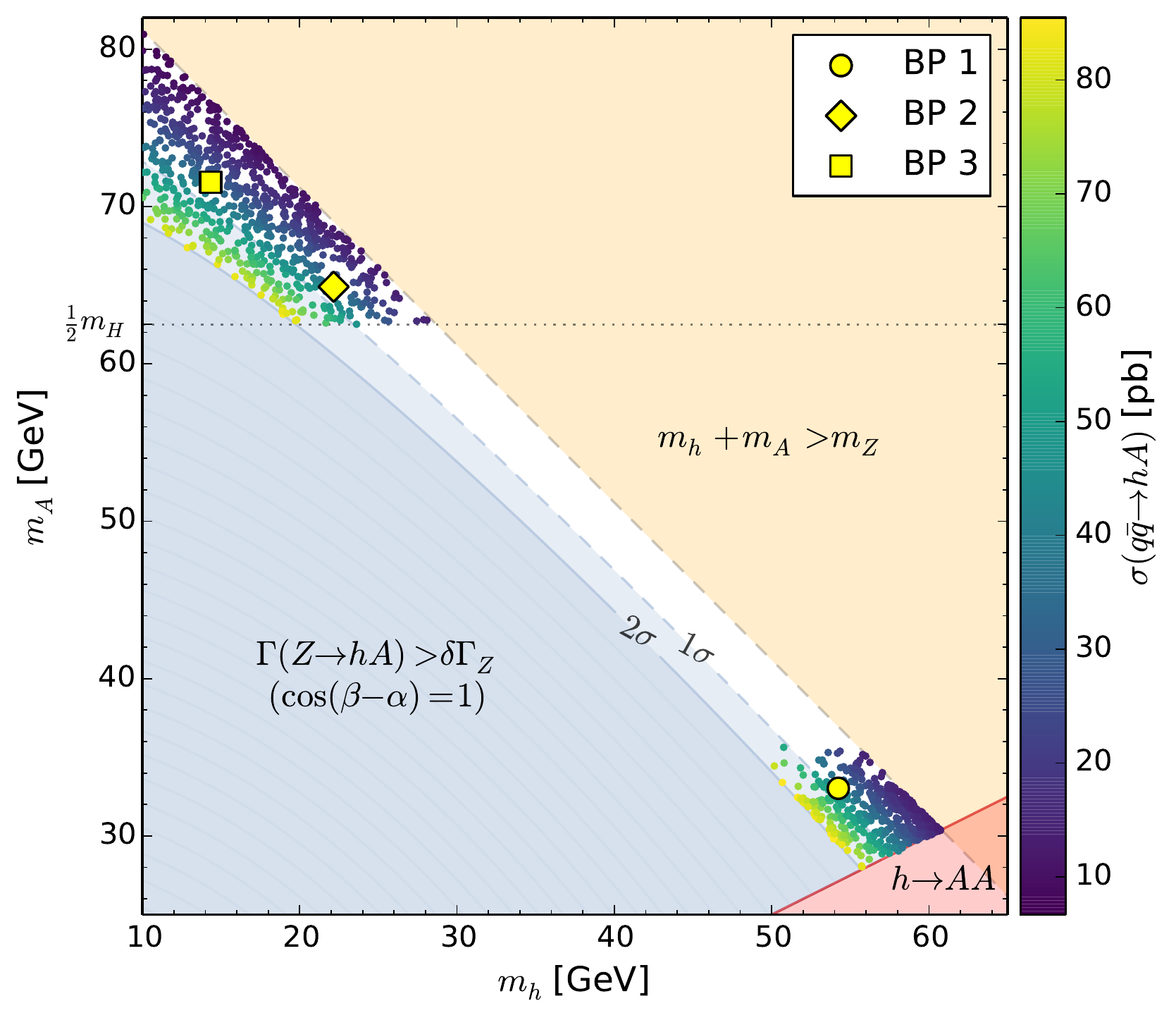}
}
\hspace{0.2cm}%
\subfloat[]{%
\includegraphics*[width=7.8cm]{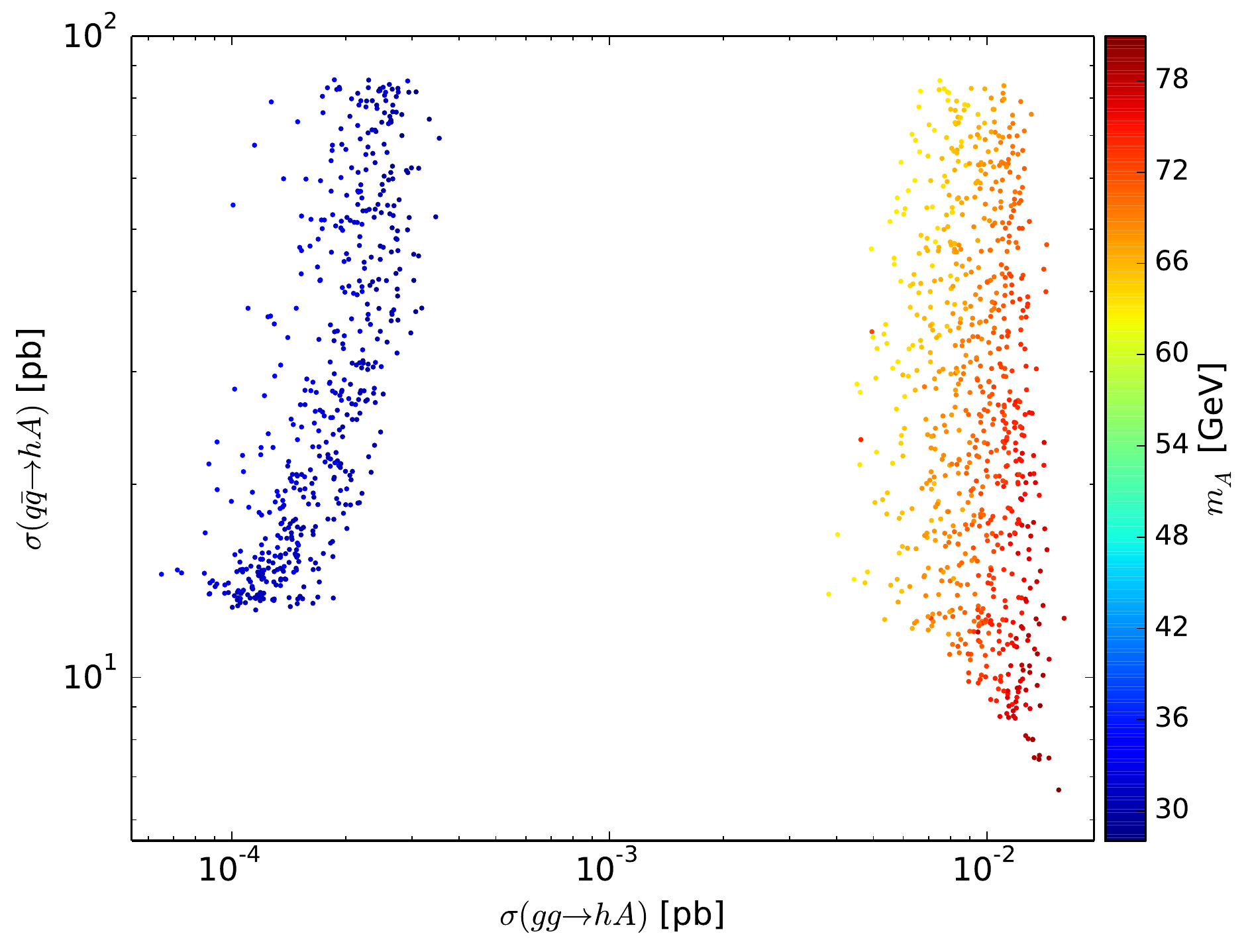}
}
\caption{(a) Successful scan points with $\Gamma(Z \to hA)$ lying within the $\delta\Gamma_Z$ at the $1\sigma$ (lighter) and $2\sigma$ (darker) levels. The color map corresponds to the total cross section for the $q\bar{q}\to hA$ process and the three BPs have been highlighted in yellow. (b) QCD vs. EW production cross sections of the $hA$ pairs, with the color map showing the mass of $A$. Taken from [7].}
\label{fig:2hdm}
\end{figure}

In figure \ref{fig:2hdm}(a) we show the good points from the scan for which the $\Gamma(Z \to hA)$ additionally lies within the 2$\sigma$ error on the experimental measurement of the total width of the $Z$ boson, $\Gamma_Z = 2.4952\pm 0.0023$\ GeV [14]. The points highlighted in yellow are the three benchmark points selected for further investigation. The color map in the figure shows the production cross section for the $q\bar{q}\to hA$ process at the LHC with $\sqrt{s} = 13$\ TeV, calculated using MadGraph5\_aMC@NLO, which evidently grows as $m_h+m_A$ gets smaller. Near the top left corner of the figure $m_A > m_h$, and we see a high density of points. The points disappear when the $H \to AA$ decay channel opens up (for $m_A < m_H/2$) , potentially leading to a significant reduction in the signal strengths of $H$ in the SM final states. The points start reappearing for $m_A < 35$\ GeV, near the bottom right corner of the figure, when the $HAA$ decay gets sufficiently suppressed. But they disappear again for $m_A < m_h/2$, where the $h \to AA$ decay channel, severely constrained by the LEP searches, is kinematically available.

Figure \ref{fig:2hdm}(b) shows that the $q\bar{q}\to hA$ production cross section at the 13\ TeV LHC can exceed the $gg\to hA$ one, calculated using [31], by a few orders of magnitude, reaching up to about 90 pb. In table \ref{tab:BP} we list the cross sections corresponding to the two production modes for the three BPs for this model. The difference between the two cross sections is much more pronounced for the BP1, wherein $m_A < m_h$, compared to that for BP2 and BP3 with $m_h < m_A$. One can also note from the table that for BP1, $Z^*A$ is the primary decay
channel of $h$, with the dominant mode for the subsequent decay of the $A$
being the $b\bar{b}$ pair. Thus $Z^*b\bar{b}b\bar{b}$, $Z^*b\bar{b}\tau^+\tau^-$ and $Z^*\tau^+\tau^-\tau^+\tau^-$ could be the main signatures of interest. Similarly, for BPs 2 and 3 $Z^*h$ is the prominent decay mode of $A$, so that the most common final states remain the same generally. For BP3 though, the highly fermiophobic $h$ (owing to $\sin(\beta - \alpha)\to 0$) has a large BR into two photons, which could make the $Z^*\gamma\gamma\gamma\gamma$ final state an important unconventional probe of this scenario in the 2HDM-I.

\begin{table}[h!]
\begin{center}
   \caption{Cross sections (in pb) for the $gg$- and $q\bar{q}$-initiated $hA$ pair-production, corresponding to the three BPs. Also given are the leading BRs of $h$ and $A$ for each BP.}
	\begin{tabular}{lllllll}%
        \hline 
		BP & $m_h$  & $m_A$ & $\sigma (q\bar{q})$ & $\sigma (gg)$ &BR$(h\to Z^*A,\,b\bar{b},\,\gamma\gamma,\,\tau\tau)$ & BR$(A\to Z^*h,\,b\bar{b},\,\tau\tau)$ \\\hline
1 & 54.2 & 33.0 & 41.2 & $1.5 \times
                                                        10^{-4}$ &
                                                                   0.94,\,0.05,\,$<0.01$,\,$<0.01$  & 0,\,0.86,\,0.07 \\
2 & 22.2 & 64.9 & 34.4 & $7.2 \times
                                                         10^{-3}$&
                                                                   0,\,0.83,\,0.03,\,0.07 & 0.86,\,0.12,\,0.01  \\ 
3 & 14.3 & 71.6 & 31.6 & $1.1 \times
                                                        10^{-2}$ & 0,\,0.60,\,0.24,\,0.07 & 0.90,\,0.08,\,0.01 \\
        \hline 
	\end{tabular}
		\label{tab:BP}
\end{center}
\end{table}

\ack

The author would like to thank his collaborators, Nils-Erik Bomark, Rikard Enberg, Chengcheng Han, Doyoun Kim, William Klemm, Stefano Moretti, Myeonghun Park and Leszek Roszkowski, each of whom were involved in one (or more) of the analyses reviewed in this contribution.

\end{document}